# Geodesic equations and their numerical solution in Cartesian coordinates on a triaxial ellipsoid

G. Panou and R. Korakitis

Department of Surveying Engineering, National Technical University of Athens, Zografou Campus, 15780 Athens, Greece

**Abstract:** In this work, the geodesic equations and their numerical solution in Cartesian coordinates on an oblate spheroid, presented by Panou and Korakitis (2017), are generalized on a triaxial ellipsoid. A new exact analytical method and a new numerical method of converting Cartesian to ellipsoidal coordinates of a point on a triaxial ellipsoid are presented. An extensive test set for the coordinate conversion is used, in order to evaluate the performance of the two methods. The direct geodesic problem on a triaxial ellipsoid is described as an initial value problem and is solved numerically in Cartesian coordinates. The solution provides the Cartesian coordinates and the angle between the line of constant $\lambda$ and the geodesic, at any point along the geodesic. Also, the Liouville constant is computed at any point along the geodesic, allowing to check the precision of the method. An extensive data set of geodesics is used, in order to demonstrate the validity of the numerical method for the geodesic problem. We conclude that a complete, stable and precise solution of the problem is accomplished.

## 1. Introduction

It is known that a triaxial ellipsoid is used as a model in geodesy and other interdisciplinary sciences, even in medicine. For example, it is used as a geometrical and physical model of the Earth and other celestial objects. Also, it is used as a geometrical model of the cornea and retina of the human eye (Aguirre 2018). Other applications of a triaxial ellipsoid are mentioned in Panou et al. (2016).

In order to describe a problem using a triaxial ellipsoid as model, it is necessary to introduce a triaxial coordinate system (see Panou 2014, Panou et al. 2016). In many applications the ellipsoidal coordinate system is used, which is a triply orthogonal system. Comments on the variants of the ellipsoidal coordinates are presented in Panou (2014). It is important that the ellipsoidal coordinates constitute an orthogonal net of curves on the triaxial ellipsoid.

In this work, the general exact analytical method of converting the Cartesian coordinates to the ellipsoidal coordinates, presented by Panou (2014), is specified for points exclusively on the surface of a triaxial ellipsoid. Another exact analytical method is described in Baillard (2013). Furthermore, a new numerical method of converting the Cartesian coordinates $(x, y, z)$ to ellipsoidal coordinates $(\beta, \lambda)$, which is based on the method of least squares, is presented. We note that another numerical method is developed by Bektaş (2015), which is also presented in Florinsky (2018). The precision of the exact analytical methods, which involve complex expressions, suffer when one approaches singular points and/or when executed on a computer with limited precision. On the other hand, numerical methods, which essentially involve iterative approximations, can be more precise but the execution time, difficult to predict, may be longer.

Traditionally, there are two problems concerning geodesics on a triaxial ellipsoid: (i) the direct problem: given a point $\Sigma_0$ on a triaxial ellipsoid, together with a direction $\alpha_0$ and the geodesic distance $s_{01}$ to a point $\Sigma_1$, determine the point $\Sigma_1$ and the direction $\alpha_1$ at this point, and (ii) the inverse problem: given two points $\Sigma_0$ and $\Sigma_1$ on a triaxial ellipsoid, determine the geodesic distance $s_{01}$ between them and the directions $\alpha_0$, $\alpha_1$ at the end points. These problems have a long history, as reviewed by Karney (2018a).

There are several methods of solving the above two problems. In general, the methods make use of the elliptic integrals presented by Jacobi (1839), where the integrands are expressed in a variant of the ellipsoidal coordinates, e.g. Bespalov (1980), Klingenberg (1982), Baillard (2013), Karney (2018b) and include a constant presented by Liouville (1844). On the other hand, there are methods which make use of the differential equations of the geodesics on a triaxial ellipsoid, e.g. Holmstrom (1976), Knill and Teodorescu (2009) and Panou (2013). Finally, Shebl and Farag (2007) use the technique of conformal mapping in order to approximate a geodesic on a triaxial ellipsoid. Because the elliptic integrals of the classical work of Jacobi (1839) have singularities, the methods which use them are preferable in the study of the qualitative characteristics of the geodesics, as presented in Arnold (1989), together with excellent illustrations by Karney (2018b). On the other hand, differential equations of the geodesics can be directly solved using an approximate analytical method (Holmstrom 1976) or a numerical method (Knill and Teodorescu 2009). It is worth emphasizing that, as presented in Panou and Korakitis (2017), geodesic equations expressed in Cartesian coordinates are insensitive to singularities. Although Holmstrom (1976) expressed the geodesic equations on a triaxial ellipsoid in Cartesian coordinates, his approximate analytical solution is of low precision.

In this work, the geodesic equations and their numerical solution in Cartesian coordinates on a triaxial ellipsoid are presented. Since the numerical solution involves computations at many points along the geodesic, it can be used as a convenient and efficient approach to trace the full path of the geodesic. Also, part of this solution constitutes the solution of the direct geodesic problem. Furthermore, in contrast to Holmstrom (1976), we make use of the ellipsoidal coordinates which are involved in Liouville equation, allowing to check the precision of the method.

## 2. Ellipsoidal to Cartesian coordinates conversion and vice versa

### 2.1. From ellipsoidal to Cartesian coordinates

A triaxial ellipsoid in Cartesian coordinates is described by

$$\frac{x^2}{a_x^2} + \frac{y^2}{a_y^2} + \frac{z^2}{b^2} = 1, \quad 0 < b < a_y < a_x \tag{1}$$

where $a_x$, $a_y$ and $b$ are its three semi-axes. The linear eccentricities are given by

$$E_x = \sqrt{a_x^2 - b^2}, \quad E_y = \sqrt{a_y^2 - b^2}, \quad E_e = \sqrt{a_x^2 - a_y^2} \tag{2}$$

with $E_e^2 = E_x^2 - E_y^2$. The Cartesian coordinates $(x, y, z)$ of a point on the triaxial ellipsoid can be obtained from the ellipsoidal coordinates $(\beta, \lambda)$ by the following expressions (Jacobi 1839)

$$x = \frac{a_x}{E_x} B^{1/2} \cos\lambda \tag{3}$$

$$y = a_y \cos\beta \sin\lambda \tag{4}$$

$$z = \frac{b}{E_x} \sin\beta L^{1/2} \tag{5}$$

where

$$B = E_x^2 \cos^2\beta + E_e^2 \sin^2\beta \tag{6}$$

and

$$L = E_x^2 - E_e^2 \cos^2\lambda \tag{7}$$

while $-\pi/2 \leq \beta \leq +\pi/2$ and $-\pi < \lambda \leq +\pi$. At the umbilical points, i.e. when $\beta = \pm\pi/2$ and $\lambda = 0$ or $\lambda = +\pi$, from Eqs. (3)-(5) we get the Cartesian coordinates $x = \pm a_x \frac{E_e}{E_x}$, $y = 0$, $z = \pm b \frac{E_y}{E_x}$. Further details on the ellipsoidal coordinates, along with their geometrical interpretation, are presented in Panou (2014). Finally, in the case of an oblate spheroid, where $a_x = a_y \equiv a$, i.e. $E_x = E_y \equiv E$ and $E_e = 0$, Eqs. (3)-(5) reduce to well-known expressions (see Heiskanen and Moritz 1967).

### 2.2. From Cartesian to ellipsoidal coordinates

#### 2.2.1. Exact analytical method

The ellipsoidal coordinates $(\beta, \lambda)$ can be obtained from the Cartesian coordinates $(x, y, z)$ of a point on the triaxial ellipsoid by solving the following quadratic equation in $t$ (see Panou 2014)

$$t^2 + c_1 t + c_0 = 0 \tag{8}$$

where

$$c_1 = x^2 + y^2 + z^2 - (a_x^2 + a_y^2 + b^2) \tag{9}$$

and

$$c_0 = a_x^2 a_y^2 + a_x^2 b^2 + a_y^2 b^2 - (a_y^2 + b^2)x^2 - (a_x^2 + b^2)y^2 - (a_x^2 + a_y^2)z^2 \tag{10}$$

with two real roots, which can be expressed as

$$t_1 = c_0/t_2 \tag{11}$$

and

$$t_2 = \left(-c_1 + \sqrt{c_1^2 - 4c_0}\right)/2 \tag{12}$$

The connection between the roots $t_1$, $t_2$ and the ellipsoidal coordinates $(\beta, \lambda)$ is given by the relations

$$t_1 = a_y^2 \sin^2\beta + b^2 \cos^2\beta \tag{13}$$

$$t_2 = a_x^2 \sin^2\lambda + a_y^2 \cos^2\lambda \tag{14}$$

where $b^2 \leq t_1 \leq a_y^2$ and $a_y^2 \leq t_2 \leq a_x^2$, while $t_1 = t_2 = a_y^2$ at the umbilical points. Inverting Eqs. (13) and (14) results in

$$\beta = \arctan\left(\sqrt{\frac{t_1 - b^2}{a_y^2 - t_1}}\right) = \operatorname{arccot}\left(\sqrt{\frac{a_y^2 - t_1}{t_1 - b^2}}\right) \tag{15}$$

$$\lambda = \arctan\left(\sqrt{\frac{t_2 - a_y^2}{a_x^2 - t_2}}\right) = \operatorname{arccot}\left(\sqrt{\frac{a_x^2 - t_2}{t_2 - a_y^2}}\right) \tag{16}$$

where the conventions with regard to the proper quadrant for the $\beta$ and $\lambda$ need to be applied from the signs of $x$, $y$ and $z$. In the case of an oblate spheroid, corresponding expressions have been presented in Heiskanen and Moritz (1967).

### 2.2.2. Numerical method

Assuming that the Cartesian coordinates $x$, $y$ and $z$ of a point on the triaxial ellipsoid are measurements, the method of least squares (Ghilani and Wolf 2006) may be employed to obtain the best estimates of the ellipsoidal coordinates $\beta$ and $\lambda$. This technique requires writing Eqs. (3)-(5) in the form

$$\frac{a_x}{E_x} B^{1/2} \cos\lambda = x + v_1 \tag{17}$$

$$a_y \cos\beta \sin\lambda = y + v_2 \tag{18}$$

$$\frac{b}{E_x} \sin\beta L^{1/2} = z + v_3 \tag{19}$$

which are non-linear equations and hence the solution process is iterative. This means that approximate values of ellipsoidal coordinates are assumed, corrections are computed and the approximate values are updated. The process is repeated until the corrections become negligible.

The linear approximation of Eqs. (17)-(19) can be represented in matrix form as

$$J \begin{bmatrix} \delta\beta \\ \delta\lambda \end{bmatrix} = \boldsymbol{\delta\ell} + \boldsymbol{v} \tag{20}$$

where

$$\mathbf{J} = \begin{bmatrix} \frac{\partial x}{\partial \beta} & \frac{\partial x}{\partial \lambda} \\ \frac{\partial y}{\partial \beta} & \frac{\partial y}{\partial \lambda} \\ \frac{\partial z}{\partial \beta} & \frac{\partial z}{\partial \lambda} \end{bmatrix} \quad (21)$$

is a 3 × 2 matrix containing the partial derivatives (Jacobian matrix)

$$\frac{\partial x}{\partial \beta} = -\frac{a_x E_y^2}{2E_x} \frac{\sin(2\beta)}{B^{1/2}} \cos\lambda \quad (22)$$

$$\frac{\partial y}{\partial \beta} = -a_y \sin\beta \sin\lambda \quad (23)$$

$$\frac{\partial z}{\partial \beta} = \frac{b}{E_x} \cos\beta L^{1/2} \quad (24)$$

$$\frac{\partial x}{\partial \lambda} = -\frac{a_x}{E_x} B^{1/2} \sin\lambda \quad (25)$$

$$\frac{\partial y}{\partial \lambda} = a_y \cos\beta \cos\lambda \quad (26)$$

$$\frac{\partial z}{\partial \lambda} = \frac{bE_e^2}{2E_x} \sin\beta \frac{\sin(2\lambda)}{L^{1/2}} \quad (27)$$

computed from the approximate ellipsoidal coordinates $\beta^0$ and $\lambda^0$,

$$\boldsymbol{\delta\ell} = \begin{bmatrix} x - x^0 \\ y - y^0 \\ z - z^0 \end{bmatrix} \quad (28)$$

is a 3 × 1 vector of terms which are "given Cartesian coordinates – computed Cartesian coordinates from the approximate ellipsoidal coordinates using Eqs. (3)-(5)" and $\boldsymbol{\upsilon}$ is a 3 × 1 vector of residuals. The corrections to the approximate ellipsoidal coordinates are the elements of the solution vector

$$\begin{bmatrix} \delta\beta \\ \delta\lambda \end{bmatrix} = \mathbf{N}^{-1} \mathbf{J}^T \boldsymbol{\delta\ell} \quad (29)$$

where

$$\mathbf{N} = \mathbf{J}^T \mathbf{J} = \begin{bmatrix} n_{11} & n_{12} \\ n_{21} & n_{22} \end{bmatrix} \quad (30)$$

and hence

$$\mathbf{N}^{-1} = \frac{1}{n_{11}n_{22} - n_{12}n_{21}} \begin{bmatrix} n_{22} & -n_{12} \\ -n_{21} & n_{11} \end{bmatrix} \quad (31)$$

One should note that the determinant of matrix **N** ($n_{11}n_{22} - n_{12}n_{21}$) equals zero at the umbilical points. The updated values of the approximate ellipsoidal coordinates are

$$\begin{bmatrix} \beta \\ \lambda \end{bmatrix} = \begin{bmatrix} \beta^0 \\ \lambda^0 \end{bmatrix} + \begin{bmatrix} \delta\beta \\ \delta\lambda \end{bmatrix} \tag{32}$$

The residuals are computed from Eq. (20) and an estimate of the variance factor $\hat{\sigma}_0^2$ can be computed using the following equation

$$\hat{\sigma}_0^2 = \mathbf{v}^T \mathbf{v} \tag{33}$$

The iterative process is terminated when the corrections $\delta\beta$ and $\delta\lambda$ become negligible. Another criterion of ending the iterative process is the convergence of the variance factor $\hat{\sigma}_0^2$ which, in the case of measurements of equal precision, is an estimate of the variance of the Cartesian coordinates computed from the adjustment (a posteriori). Finally, the variance-covariance matrix of the computed ellipsoidal coordinates is given by

$$\hat{\mathbf{V}}_{\begin{bmatrix} \beta \\ \lambda \end{bmatrix}} = \hat{\sigma}_0^2 \mathbf{N}^{-1} \tag{34}$$

Comparing the previous two methods, we note that the operations in the exact analytical method lead to a loss of accuracy for points near the planes $x = 0$, $y = 0$ and $z = 0$. Also, the importance of the numerical method is that we avoid the degeneracy of the variable $t_2$ in Eq. (16), which may yield inaccurate results, since the intervals of variation of the coordinates $\beta$ and $\lambda$ remain invariants as a triaxial ellipsoid transforms to an oblate spheroid, where $t_2 = a^2$. Therefore, the values resulting from the exact analytical method can be considered as initial, approximate ellipsoidal coordinates in the numerical method. However, because Eqs. (3)-(5) are numerically stable, the results of both methods can be checked by comparing the resulting Cartesian coordinates ($x^0, y^0, z^0$) with the given Cartesian coordinates ($x, y, z$), e.g. by the simple formula

$$\delta r = \sqrt{(x - x^0)^2 + (y - y^0)^2 + (z - z^0)^2} \tag{35}$$

## 3. Geodesic equations

The geodesic initial value problem, expressed in Cartesian coordinates on a triaxial ellipsoid, consists of determining a geodesic, parametrized by its arc length $s$, $x = x(s), y = y(s), z = z(s)$, with angles $\alpha = \alpha(s)$ along it, between the line of constant $\lambda$ and the geodesic, which passes through a given point $\Sigma_0(x(0), y(0), z(0))$ in a known direction (given angle $\alpha_0 = \alpha(0)$) and has a certain length $s_{01}$.

Now, we consider a triaxial ellipsoid which is described in Cartesian coordinates ($x, y, z$) by

$$S(x, y, z) \doteq x^2 + \frac{y^2}{1-e_e^2} + \frac{z^2}{1-e_x^2} - a_x^2 = 0 \tag{36}$$

where the squared eccentricities $e_x^2$ and $e_e^2$ are given by

$$e_x^2 = (a_x^2 - b^2)/a_x^2, \quad e_e^2 = (a_x^2 - a_y^2)/a_x^2 \tag{37}$$

It is well-known, from the theory of differential geometry, that the principal normal to the geodesic must coincide with the normal to the triaxial ellipsoid (Struik 1961), i.e.

$$\frac{d^2x/ds^2}{\partial S/\partial x} = \frac{d^2y/ds^2}{\partial S/\partial y} = \frac{d^2z/ds^2}{\partial S/\partial z} = -m \tag{38}$$

where $m$ is a function of $s$. From these equations, together with Eq. (36), it is possible to determine $x(s)$, $y(s)$, $z(s)$ and $m(s)$. Using Eq. (36), Eqs. (38) become

$$\frac{1}{x}\frac{d^2x}{ds^2} = \frac{1-e_e^2}{y}\frac{d^2y}{ds^2} = \frac{1-e_x^2}{z}\frac{d^2z}{ds^2} = -2m \tag{39}$$

Differentiating Eq. (36), we have

$$x\frac{dx}{ds} + \frac{y}{1-e_e^2}\frac{dy}{ds} + \frac{z}{1-e_x^2}\frac{dz}{ds} = 0 \tag{40}$$

and a further differentiation yields

$$x\frac{d^2x}{ds^2} + \frac{y}{1-e_e^2}\frac{d^2y}{ds^2} + \frac{z}{1-e_x^2}\frac{d^2z}{ds^2} = -\left[\left(\frac{dx}{ds}\right)^2 + \frac{1}{1-e_e^2}\left(\frac{dy}{ds}\right)^2 + \frac{1}{1-e_x^2}\left(\frac{dz}{ds}\right)^2\right] \tag{41}$$

Hence, from Eqs. (39) and (41), we obtain

$$m = \frac{h}{2H} \tag{42}$$

where

$$H = x^2 + \frac{y^2}{(1-e_e^2)^2} + \frac{z^2}{(1-e_x^2)^2} \tag{43}$$

and

$$h = \left(\frac{dx}{ds}\right)^2 + \frac{1}{1-e_e^2}\left(\frac{dy}{ds}\right)^2 + \frac{1}{1-e_x^2}\left(\frac{dz}{ds}\right)^2 \tag{44}$$

Substituting Eq. (42) into Eqs. (39), we obtain the geodesic equations in Cartesian coordinates on a triaxial ellipsoid

$$\frac{d^2x}{ds^2} + \frac{h}{H}x = 0 \tag{45}$$

$$\frac{d^2y}{ds^2} + \frac{h}{H}\frac{y}{1-e_e^2} = 0 \tag{46}$$

$$\frac{d^2z}{ds^2} + \frac{h}{H}\frac{z}{1-e_x^2} = 0 \tag{47}$$

which are subject to the initial conditions

$$x_0 = x(0), \quad \left.\frac{dx}{ds}\right|_0 = \frac{dx}{ds}(0) \tag{48}$$

$$y_0 = y(0), \quad \left.\frac{dy}{ds}\right|_0 = \frac{dy}{ds}(0) \tag{49}$$

$$z_0 = z(0), \quad \left.\frac{dz}{ds}\right|_0 = \frac{dz}{ds}(0) \tag{50}$$

where expressions for the values of the derivatives at point $\Sigma_0(x_0, y_0, z_0)$ are produced below. Hence, the direct geodesic problem is described as an initial value problem in Cartesian coordinates on a triaxial ellipsoid by Eqs. (45) to (50).

## 4. Numerical solution

In order to solve the above problem, the system of three non-linear second order ordinary differential equations (Eqs. (45) to (47)) is rewritten as a system of six first-order differential equations:

$$\frac{d}{ds}(x) = \frac{dx}{ds} \tag{51}$$

$$\frac{d}{ds}\left(\frac{dx}{ds}\right) = -\frac{h}{H}x \tag{52}$$

$$\frac{d}{ds}(y) = \frac{dy}{ds} \tag{53}$$

$$\frac{d}{ds}\left(\frac{dy}{ds}\right) = -\frac{h}{H}\frac{y}{1-e_e^2} \tag{54}$$

$$\frac{d}{ds}(z) = \frac{dz}{ds} \tag{55}$$

$$\frac{d}{ds}\left(\frac{dz}{ds}\right) = -\frac{h}{H}\frac{z}{1-e_x^2} \tag{56}$$

This system can be integrated on the interval $[0, s]$ using a numerical method, such as Runge-Kutta (see Butcher 1987). The step size $\delta s$ is given by $\delta s = s/n$, where $n$ is the number of steps. For the variables $x$, $y$ and $z$, the initial conditions are $x_0$, $y_0$ and $z_0$, respectively. To obtain the required derivatives, we proceed to describe the unit vectors to a geodesic through a point $\Sigma(x, y, z)$ on a triaxial ellipsoid (see Fig. 1).

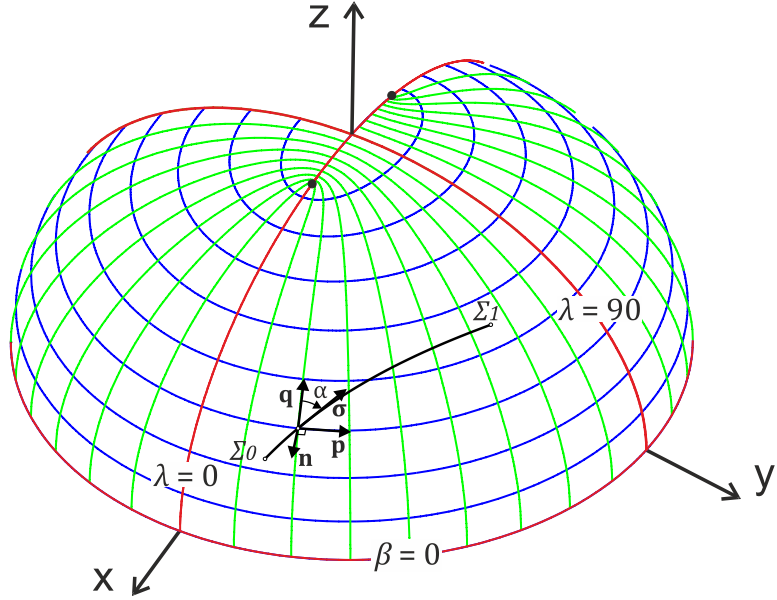

**Figure 1:** Unit vectors to a geodesic through a point $\Sigma$ on a triaxial ellipsoid: **σ** tangent to the geodesic, **n** normal to the triaxial ellipsoid, **p** tangent to the line of constant $\beta$, **q** tangent to the line of constant $\lambda$.

Let **σ** be a unit vector tangent to an arbitrary geodesic through $\Sigma$. Then, we can express **σ** in terms of the unit vectors **p**, **q** and the angle $\alpha$ between the line of constant $\lambda$ and the geodesic (Fig. 1):

$$\boldsymbol{\sigma} = \left(\frac{dx}{ds}, \frac{dy}{ds}, \frac{dz}{ds}\right) = \mathbf{p}\sin\alpha + \mathbf{q}\cos\alpha \tag{57}$$

The unit vector normal to a triaxial ellipsoid (using the gradient operator and Eqs. (36), (43)) can be expressed as (Fig. 1):

$$\mathbf{n} = (n_1, n_2, n_3) = \left(\frac{x}{H^{1/2}}, \frac{y}{(1-e_e^2)H^{1/2}}, \frac{z}{(1-e_x^2)H^{1/2}}\right) \tag{58}$$

The unit vector $\mathbf{p} = (p_1, p_2, p_3)$, tangent to the line of constant $\beta$, can be determined using Eqs. (25)-(27) and Eqs. (7) and (14) (Fig. 1):

$$p_1 = -\left(\frac{L}{Ft_2}\right)^{1/2} \frac{a_x}{E_x} B^{1/2} \sin\lambda \tag{59}$$

$$p_2 = \left(\frac{L}{Ft_2}\right)^{1/2} a_y \cos\beta \cos\lambda \tag{60}$$

$$p_3 = \frac{1}{(Ft_2)^{1/2}} \frac{bE_e^2}{2E_x} \sin\beta \sin(2\lambda) \tag{61}$$

where

$$F = E_y^2 \cos^2\beta + E_e^2 \sin^2\lambda \tag{62}$$

Also, this vector can be expressed in terms of Cartesian coordinates with the help of Eqs. (13) and (14):

$$p_1 = -sgn(y)\left(\frac{L}{Ft_2}\right)^{1/2}\frac{a_x}{E_x E_e}B^{1/2}\sqrt{t_2 - a_y^2} \tag{63}$$

$$p_2 = sgn(x)\left(\frac{L}{Ft_2}\right)^{1/2}\frac{a_y}{E_y E_e}\sqrt{(a_y^2 - t_1)(a_x^2 - t_2)} \tag{64}$$

$$p_3 = sgn(x)sgn(y)sgn(z)\frac{1}{(Ft_2)^{1/2}}\frac{b}{E_x E_y}\sqrt{(t_1 - b^2)(t_2 - a_y^2)(a_x^2 - t_2)} \tag{65}$$

where

$$B = \frac{E_x^2}{E_y^2}(a_y^2 - t_1) + \frac{E_e^2}{E_y^2}(t_1 - b^2) \tag{66}$$

$$L = t_2 - b^2 \tag{67}$$

and

$$F = t_2 - t_1 \tag{68}$$

while $sgn(x) = 1$ if $x > 0$, $sgn(x) = -1$ if $x < 0$ and $sgn(0) = 0$.

However, vector **p** has singularities at the umbilical points, where we can simply set $\mathbf{p} = (p_1, p_2, p_3) = (0, \pm 1, 0)$. Finally, in the case of an oblate spheroid, Eqs. (59)-(61) reduce to the expressions (Panou and Korakitis 2017)

$$\mathbf{p} = (p_1, p_2, p_3) = (-sin\lambda, cos\lambda, 0) \tag{69}$$

and Eqs. (63)-(65) can be replaced by

$$\mathbf{p} = (p_1, p_2, p_3) = \left(\frac{-y}{\sqrt{x^2+y^2}}, \frac{x}{\sqrt{x^2+y^2}}, 0\right) \tag{70}$$

The unit vector $\mathbf{q} = (q_1, q_2, q_3)$, tangent to the line of constant $\lambda$, can now be determined as the cross product of unit vectors **n** and **p**, i.e. $\mathbf{q} = \mathbf{n} \times \mathbf{p}$ (Fig. 1):

$$q_1 = n_2 p_3 - n_3 p_2 \tag{71}$$

$$q_2 = n_3 p_1 - n_1 p_3 \tag{72}$$

$$q_3 = n_1 p_2 - n_2 p_1 \tag{73}$$

Finally, substituting the vectors **p** and **q** into Eq. (57), we obtain the required values of the derivatives at point $\Sigma_0(x_0, y_0, z_0)$

$$\left.\frac{dx}{ds}\right|_0 = p_1(0)sin\alpha_0 + q_1(0)cos\alpha_0 \tag{74}$$

$$\left.\frac{dy}{ds}\right|_0 = p_2(0)sin\alpha_0 + q_2(0)cos\alpha_0 \tag{75}$$

$$\left.\frac{dz}{ds}\right|_0 = p_3(0)sin\alpha_0 + q_3(0)cos\alpha_0 \tag{76}$$

## 5. Angles and Liouville's constant

Taking the scalar product of Eq. (57) successively with **p** and **q** and dividing the resulting equations, yields the angle at which the geodesic cuts the curve of constant $\lambda$

$$\alpha = arctan\left(\frac{P}{Q}\right) = arccot\left(\frac{Q}{P}\right) \tag{77}$$

where

$$P = \mathbf{p} \cdot \boldsymbol{\sigma} = p_1\frac{dx}{ds} + p_2\frac{dy}{ds} + p_3\frac{dz}{ds} \tag{78}$$

$$Q = \mathbf{q} \cdot \boldsymbol{\sigma} = q_1\frac{dx}{ds} + q_2\frac{dy}{ds} + q_3\frac{dz}{ds} \tag{79}$$

Note that Eqs. (77) involve all the variables $x$, $dx/ds$, $y$, $dy/ds$, $z$ and $dz/ds$, which are obtained by the numerical integration.

Along a geodesic on a triaxial ellipsoid, the Liouville equation holds (Liouville 1844)

$$E_y^2 cos^2\beta sin^2\alpha - E_e^2 sin^2\lambda cos^2\alpha = c \tag{80}$$

where $c$ is the Liouville constant. Also, this equation can be expressed in terms of Cartesian coordinates with the help of Eqs. (13) and (14):

$$a_y^2 - (t_1 sin^2\alpha + t_2 cos^2\alpha) = c \tag{81}$$

At any value of the independent variable $s$, we can estimate the difference $\delta c = c - c_0$ between the computed value $c$ and the known value $c_0$ at point $\Sigma_0$, from the given $\beta_0$, $\lambda_0$ and $\alpha_0$, by means of Liouville's equation (Eq. (80)). Furthermore, because the numerical integration is performed in space, we can compute, at any value of $s$, the function $S$, given by Eq. (36). Therefore, we can check both the precision of the method and of the numerical integration, since the difference $\delta c$ and the function $S$ should be zero (meters squared) at any point along the geodesic on a triaxial ellipsoid.

## 6. Numerical experiments
### 6.1. Test set for coordinates conversion

In order to validate the two methods of conversion presented above and to evaluate their performance, we used an extensive test set of points. This is a set of 1725 points on a triaxial ellipsoid, distributed into ten groups, as described in Table 1, where N stands for the number of points in each group. For simplicity and without loss of generality, $\beta$ and $\lambda$ were chosen in $[0°, 90°]$.

Table 1: Description of the points in the test set

| Group | $\beta$ | $\lambda$ | Case | N |
|---|---|---|---|---|
| 1 | $5° - 85°$ every $5°$ | $5° - 85°$ every $5°$ | 1st octant | 289 |
| 2 | $0°$ | $0° - 90°$ every $5°$ | $xy$ – plane | 19 |
| 3 | $5° - 85°$ every $5°$ | $0°$ | $xz$ – plane | 17 |
| 4 | $90°$ | $5° - 90°$ every $5$ | | 18 |
| 5 | $89.0° - 89.9 \ldots 9°$ up to 14 decimals | $1.0° - 0.0 \ldots 1°$ up to 14 decimals | near umbilic | 225 |
| 6 | $5° - 85°$ every $5°$ | $90°$ | $yz$ – plane | 17 |
| 7 | $1.0° - 0.0 \ldots 1°$ up to 14 decimals | $0° - 90°$ every $5°$ | near $xy$ – plane | 285 |
| 8 | $0° - 90°$ every $5°$ | $1.0° - 0.0 \ldots 1°$ up to 14 decimals | near $xz$ – plane | 285 |
| 9 | $89.0° - 89.9 \ldots 9°$ up to 14 decimals | $0° - 90°$ every $5°$ | | 285 |
| 10 | $0° - 90°$ every $5°$ | $89.0° - 89.9 \ldots 9°$ up to 14 decimals | near $yz$ – plane | 285 |

Using a triaxial ellipsoid with $a_x = 6378172$ m, $a_y = 6378103$ m and $b = 6356753$ m, (Ligas 2012) the Cartesian coordinates for any point were computed using Eqs. (3)-(5).

All algorithms were coded in C++, were compiled by the open-source GNU GCC compiler (at Level 2 optimization) and employing the open-source "libquadmath", the GCC Quad-Precision Math Library, which provides a precision of 33 digits. In contrast, use of the C++ long-double standard type (referred simply as double in the following sections) provides a precision of 18 digits. The codes were executed on a personal computer running a 64-bit Linux Debian operating system. The main characteristics of the hardware were: Intel Core i5-2430M CPU (clocked at 2.4 GHz) and 6 GB of RAM.

### 6.1.1. Results

The exact analytical method was applied using double and quad precision and the Cartesian coordinates $x$, $y$ and $z$ as input data. From the resulting $\beta^0$ and $\lambda^0$ at any point, we computed the differences $\delta\beta = \beta - \beta^0$ and $\delta\lambda = \lambda - \lambda^0$ and recorded the $max|\delta\beta|$ and the $max|\delta\lambda|$ for every Group. Furthermore, the results at any point were converted back to Cartesian coordinates, we computed the value $\delta r$ using Eq. (35) and recorded the $max|\delta r|$ for every Group. All results are presented in Table 2.

Table 2: Performance of the exact analytical method using double and quad precision

| Group | double | | | quad | | |
|---|---|---|---|---|---|---|
| | $max\|\delta\beta\|$ (") | $max\|\delta\lambda\|$ (") | $max\delta r$ (m) | $max\|\delta\beta\|$ (") | $max\|\delta\lambda\|$ (") | $max\delta r$ (m) |
| 1 | $1.27 \cdot 10^{-6}$ | $1.38 \cdot 10^{-4}$ | $4.13 \cdot 10^{-4}$ | $5.14 \cdot 10^{-21}$ | $1.04 \cdot 10^{-18}$ | $2.80 \cdot 10^{-18}$ |
| 2 | $1.66 \cdot 10^{-2}$ | $1.54 \cdot 10^{-1}$ | $4.78$ | $1.05 \cdot 10^{-9}$ | $9.45 \cdot 10^{-21}$ | $3.23 \cdot 10^{-8}$ |

| 3 | $5.40 \cdot 10^{-7}$ | 2.44 | 8.44 | $4.62 \cdot 10^{-21}$ | $6.44 \cdot 10^{-8}$ | $6.55 \cdot 10^{-7}$ |
|---|---|---|---|---|---|---|
| 4 | $9.32 \cdot 10^{-1}$ | $1.46 \cdot 10^{-2}$ | 6.44 | $1.41 \cdot 10^{-7}$ | $1.72 \cdot 10^{-16}$ | $6.33 \cdot 10^{-7}$ |
| 5 | 50.6 | $8.90 \cdot 10^{2}$ | 8.08 | $1.63 \cdot 10^{-2}$ | $2.86 \cdot 10^{-1}$ | $6.97 \cdot 10^{-7}$ |
| 6 | $1.24 \cdot 10^{-7}$ | $5.24 \cdot 10^{-1}$ | 8.33 | $3.27 \cdot 10^{-21}$ | $1.04 \cdot 10^{-7}$ | $5.87 \cdot 10^{-7}$ |
| 7 | $1.81 \cdot 10^{-2}$ | $3.10 \cdot 10^{-1}$ | 9.59 | $1.05 \cdot 10^{-9}$ | $1.89 \cdot 10^{-8}$ | $5.85 \cdot 10^{-7}$ |
| 8 | 50.6 | $8.90 \cdot 10^{2}$ | 10.6 | $1.55 \cdot 10^{-2}$ | $2.50 \cdot 10^{-1}$ | $7.11 \cdot 10^{-7}$ |
| 9 | 50.6 | $8.90 \cdot 10^{2}$ | 10.8 | $1.63 \cdot 10^{-2}$ | $2.86 \cdot 10^{-1}$ | $6.97 \cdot 10^{-7}$ |
| 10 | $2.01 \cdot 10^{-1}$ | 3.56 | 10.4 | $2.16 \cdot 10^{-8}$ | $1.97 \cdot 10^{-7}$ | $6.68 \cdot 10^{-7}$ |

Comparing the results of $max|\delta r|$ presented in Table 2 between the double and quad precision, we conclude that only quad precision provides results suitable for most practical applications. Also, we remark that only the quantity $\delta r$ can be computed in a problem starting with knowledge of Cartesian coordinates only.

Similarly, the numerical method was applied using double and quad precision and the Cartesian coordinates $x$, $y$ and $z$ as input data. From the resulting $\beta^0$ and $\lambda^0$ at any point, we computed the differences $\delta\beta = \beta - \beta^0$ and $\delta\lambda = \lambda - \lambda^0$ and recorded the $max|\delta\beta|$ and the $max|\delta\lambda|$ for every Group. It is known that, using the method of least squares, we compute the matrix **N** and hence the variance-covariance matrix of the computed values of $\beta$ and $\lambda$. Therefore, we can estimate the errors of $\beta$ and $\lambda$ (from the diagonal elements of matrix $\widehat{\mathbf{V}}$), so we recorded the $max|\hat{\sigma}_\beta|$ and $max|\hat{\sigma}_\lambda|$ for every Group. Furthermore, we computed the value $\delta r$ using Eq. (35) and recorded the $max|\delta r|$ for every Group. Finally, we recorded the mean and the maximum value of iterations $i$, which were needed, using as criterion of convergence of the standard error $\hat{\sigma}_0$ the values $10^{-19}$ and $10^{-33}$, for double and quad precision, respectively. All results are presented in Tables 3 and 4.

Table 3: Performance of the numerical method using double precision

| Group | $max|\delta\beta|$ (") | $max|\delta\lambda|$ (") | $max|\hat{\sigma}_\beta|$ (") | $max|\hat{\sigma}_\lambda|$ (") | $max\delta r$ (m) | $mean(i)$ | $max(i)$ |
|---|---|---|---|---|---|---|---|
| 1 | $3.75 \cdot 10^{-14}$ | $6.25 \cdot 10^{-14}$ | $7.76 \cdot 10^{-9}$ | $6.46 \cdot 10^{-8}$ | $1.16 \cdot 10^{-12}$ | 3 | 5 |
| 2 | 0 | $2.50 \cdot 10^{-14}$ | $5.52 \cdot 10^{-9}$ | $5.50 \cdot 10^{-9}$ | $8.20 \cdot 10^{-13}$ | 3 | 5 |
| 3 | $1.87 \cdot 10^{-14}$ | 0 | $3.56 \cdot 10^{-9}$ | $3.42 \cdot 10^{-8}$ | $9.10 \cdot 10^{-13}$ | 3 | 4 |
| 4 | 0 | $1.58 \cdot 10^{-13}$ | $1.75 \cdot 10^{-8}$ | $3.08 \cdot 10^{-7}$ | $4.82 \cdot 10^{-13}$ | 4 | 6 |
| 5 | $3.33 \cdot 10^{-1}$ | 5.81 | $8.12 \cdot 10^{-3}$ | $1.43 \cdot 10^{-1}$ | $2.92 \cdot 10^{-4}$ | 6 | 10 |
| 6 | $2.50 \cdot 10^{-14}$ | 0 | $4.30 \cdot 10^{-9}$ | $8.24 \cdot 10^{-9}$ | $6.53 \cdot 10^{-13}$ | 3 | 5 |
| 7 | $1.56 \cdot 10^{-15}$ | $2.50 \cdot 10^{-14}$ | $6.11 \cdot 10^{-9}$ | $6.08 \cdot 10^{-9}$ | $1.02 \cdot 10^{-12}$ | 3 | 5 |
| 8 | $3.33 \cdot 10^{-1}$ | 5.81 | $1.19 \cdot 10^{-3}$ | $2.09 \cdot 10^{-2}$ | $2.92 \cdot 10^{-4}$ | 3 | 9 |
| 9 | $1.98 \cdot 10^{-1}$ | 3.48 | $8.09 \cdot 10^{-3}$ | $1.42 \cdot 10^{-1}$ | $1.03 \cdot 10^{-4}$ | 4 | 9 |
| 10 | $2.50 \cdot 10^{-14}$ | $2.50 \cdot 10^{-14}$ | $7.59 \cdot 10^{-9}$ | $5.34 \cdot 10^{-8}$ | $1.16 \cdot 10^{-12}$ | 3 | 5 |

Table 4: Performance of the numerical method using quad precision

| Group | $max|\delta\beta|$ (") | $max|\delta\lambda|$ (") | $max|\hat{\sigma}_\beta|$ (") | $max|\hat{\sigma}_\lambda|$ (") | $max\delta r$ (m) | $mean(i)$ | $max(i)$ |
|---|---|---|---|---|---|---|---|
| 1 | $2.00 \cdot 10^{-28}$ | $4.88 \cdot 10^{-28}$ | $4.66 \cdot 10^{-23}$ | $3.68 \cdot 10^{-22}$ | $6.96 \cdot 10^{-27}$ | 3 | 6 |
| 2 | 0 | $1.77 \cdot 10^{-28}$ | $4.12 \cdot 10^{-23}$ | $4.10 \cdot 10^{-23}$ | $6.06 \cdot 10^{-27}$ | 3 | 5 |
| 3 | $1.77 \cdot 10^{-28}$ | 0 | $3.63 \cdot 10^{-23}$ | $3.13 \cdot 10^{-22}$ | $5.42 \cdot 10^{-27}$ | 3 | 4 |
| 4 | 0 | $7.60 \cdot 10^{-28}$ | $2.48 \cdot 10^{-22}$ | $4.35 \cdot 10^{-21}$ | $4.85 \cdot 10^{-27}$ | 3 | 5 |
| 5 | $6.36 \cdot 10^{-5}$ | $1.12 \cdot 10^{-3}$ | $1.90 \cdot 10^{-6}$ | $3.35 \cdot 10^{-5}$ | $1.06 \cdot 10^{-11}$ | 4 | 10 |
| 6 | $2.22 \cdot 10^{-28}$ | 0 | $3.04 \cdot 10^{-23}$ | $2.58 \cdot 10^{-22}$ | $4.71 \cdot 10^{-27}$ | 3 | 5 |

| 7 | $1.80 \cdot 10^{-29}$ | $2.22 \cdot 10^{-28}$ | $3.95 \cdot 10^{-23}$ | $3.94 \cdot 10^{-23}$ | $6.51 \cdot 10^{-27}$ | 3 | 5 |
|---|---|---|---|---|---|---|---|
| 8 | $5.55 \cdot 10^{-5}$ | $9.76 \cdot 10^{-4}$ | $8.45 \cdot 10^{-11}$ | $1.48 \cdot 10^{-9}$ | $8.13 \cdot 10^{-12}$ | 3 | 9 |
| 9 | $6.83 \cdot 10^{-5}$ | $1.20 \cdot 10^{-3}$ | $3.01 \cdot 10^{-10}$ | $5.28 \cdot 10^{-9}$ | $1.23 \cdot 10^{-11}$ | 3 | 8 |
| 10 | $2.22 \cdot 10^{-28}$ | $8.87 \cdot 10^{-29}$ | $4.49 \cdot 10^{-23}$ | $5.69 \cdot 10^{-22}$ | $6.72 \cdot 10^{-27}$ | 3 | 6 |

Comparing the results of $max|\delta r|$ presented in Tables 3 and 4 between the double and quad precision, we conclude that both precisions can give results suitable for most practical applications (better than 1 mm for double and 1 nm for quad precision).

### 6.2. Data set for geodesics

In order to evaluate the performance of the presented method with respect to stability and precision, we used an extensive data set of 150000 geodesics for a triaxial ellipsoid with $a_x = 6378172$ m, $a_y = 6378103$ m and $b = 6356753$ m (Ligas 2012). The geodesics of the set were distributed into five groups (A – E) with different qualitative characteristics, as described in Tables 5 – 9. Each geodesic of the data set was defined by the values of $\beta_0$ (in degrees), $\lambda_0$ (in degrees), $\alpha_0$ (clockwise from $\lambda = $ constant in degrees) and $s_{01}$ (in meters). Furthermore, $\beta_0$, $\lambda_0$ and $\alpha_0$ were taken to be multiples of $10^{-12}$ deg and $s_{01}$ a multiple of 0.1 μm in [0 m, 20003987.55893028 m], where the upper bound for the $s_{01}$ is the geodesic distance between opposite umbilical points (i.e. the half arc length of the ellipse with axes $a_x$ and $b$).

Table 5: Description of the geodesics in the Group A

| Group A: $\beta_0 \in (0°, 90°)$,    $\lambda_0 = -90°$,    $\alpha_0 \in [0°, 180°]$ | | |
|---|---|---|
| Subgroup | Case | Number |
| A.1 | randomly distributed | 5000 |
| A.2 | nearly antipodal | 5000 |
| A.3 | short distances | 5000 |
| A.4 | $\beta_0 \cong 90°$ (one end near $z = b$) | 5000 |
| A.5 | $\beta_0 \cong 90°$ & $s_{01} \cong 20003879$ m (both ends near opposite $z = b$) | 5000 |
| A.6 | $\alpha_0 \cong 0°$ or $\alpha_0 \cong 180°$ | 5000 |
| A.7 | $\beta_0 \cong 0°$ & $\alpha_0 \cong 90°$ | 5000 |
| A.8 | $\alpha_0 \cong 90°$ | 5000 |
| A.9 | $\alpha_0 = 90°$ | 5000 |

Table 6: Description of the geodesics in the Group B

| Group B: $\beta_0 \in (0°, 90°)$,    $\lambda_0 = 0°$,    $\alpha_0 \in [0°, 180°]$,    $c_0 \geq 0$ | | |
|---|---|---|
| Subgroup | Case | Number |
| B.1 | randomly distributed | 5000 |
| B.2 | nearly antipodal | 5000 |
| B.3 | short distances | 5000 |
| B.4 | $\beta_0 \cong 90°$ (one end near an umbilic point) | 5000 |
| B.5 | $\beta_0 \cong 90°$ & $s_{01} \cong 20003988$ m (both ends near opposite umbilical points) | 5000 |
| B.6 | $\alpha_0 \cong 0°$ or $\alpha_0 \cong 180°$ | 5000 |
| B.7 | $\beta_0 \cong 0°$ & $\alpha_0 \cong 90°$ | 5000 |

| B.8 | $\alpha_0 \cong 90°$ | 5000 |
| B.9 | $\alpha_0 = 90°$ | 5000 |

Table 7: Description of the geodesics in the Group C

| Group C: $\beta_0 = 90°$, $\lambda_0 \in (0°, 90°]$, $\alpha_0 \in [90°, 270°]$, $c_0 \leq 0$ | | |
|---|---|---|
| Subgroup | Case | Number |
| C.1 | randomly distributed | 5000 |
| C.2 | nearly antipodal | 5000 |
| C.3 | short distances | 5000 |
| C.4 | $\lambda_0 \cong 90°$ (one end near $z = b$) | 5000 |
| C.5 | $\lambda_0 \cong 0°$ (one end near an umbilic point) | 5000 |
| C.6 | $\lambda_0 \cong 90°$ & $s_{01} \cong 20003879$ m (both ends near opposite $z = b$) | 5000 |
| C.7 | $\lambda_0 \cong 0°$ & $s_{01} \cong 20003988$ m (both ends near opposite umbilical points) | 5000 |
| C.8 | $\lambda_0 = 45°$ & $\alpha_0 \cong 180°$ | 5000 |
| C.9 | $\alpha_0 = 90°$ | 5000 |

Table 8: Description of the geodesics in the Group D

| Group D (Umbilical geodesics): $\beta_0 = 90°$, $\lambda_0 = 0°$, $\alpha_0 \in [0°, 180°]$, $c_0 = 0$ | | |
|---|---|---|
| Subgroup | Case | Number |
| D.1 | randomly distributed | 4639 |
| D.2 | $\alpha_0$ every 0.5°, $s_{01} = 20003987.55893028$ m | 361 |

Table 9: Description of the geodesics in the Group E

| Group E: $\beta_0 \in (-90°, 90°)$, $\lambda_0 \in (-180°, 180°)$, $\alpha_0 \in (0°, 360°)$, $s_{01} \in (1000 \text{ m}, 20003879 \text{ m})$ | |
|---|---|
| Case | Number |
| randomly distributed | 10000 |

### 6.2.1. Results

The direct geodesic problem in Cartesian coordinates was solved using the input data $\beta_0$, $\lambda_0$, $\alpha_0$ and $s_{01}$. At the starting point, the Cartesian coordinates $(x_0, y_0, z_0)$ were computed using Eqs. (3)-(5) and the Liouville constant $c_0$ using Eq. (80). For each geodesic in the data set, the unit vector **p** was computed using Eqs. (59)-(61) and the system of first-order differential equations (Eqs. (51)-(56)) was integrated using the fourth-order Runge-Kutta numerical method (see Butcher 1987) with 20000 steps. This number of steps was chosen because the effects of the number of steps for the same problem on an oblate spheroid were studied in the work of Panou and Korakitis (2017). All algorithms were coded and executed on the system described in section 6.1.

The results $x_1$, $y_1$ and $z_1$ at the end point were converted to ellipsoidal coordinates $\beta_1$ and $\lambda_1$ using the numerical method of subsection 2.2.2. Then, the unit vector **p** was computed using Eqs. (59)-(61) and the angle $\alpha_1$ from Eq. (77). Also, the Liouville constant $c_1$ was computed using Eq. (80) and the difference $\delta c_1 = c_1 - c_0$ recorded. Furthermore, we recorded the value $S_1$ of the function $S$, given by Eq. (36). Detailed results of the maximum values of $|\delta c_1|$ and $|S_1|$ for each subgroup in double and quad precision are presented in Tables 10 – 14. We emphasize that the value of $S_1$ is affected by the precision of the numerical integration (which depends on the number of steps), while the value of $\delta c_1$ is mostly affected by the errors of the conversion of Cartesian to ellipsoidal coordinates at the end point.

Table 10: Results in the Group A

| Subgroup | double | | quad | |
|---|---|---|---|---|
| | $max|\delta c_1|$ (m$^2$) | $max|S_1|$ (m$^2$) | $max|\delta c_1|$ (m$^2$) | $max|S_1|$ (m$^2$) |
| A.1 | $1.83 \cdot 10^{-6}$ | $1.21 \cdot 10^{-3}$ | $1.17 \cdot 10^{-11}$ | $5.05 \cdot 10^{-7}$ |
| A.2 | $1.59 \cdot 10^{-6}$ | $1.20 \cdot 10^{-3}$ | $6.07 \cdot 10^{-12}$ | $5.32 \cdot 10^{-7}$ |
| A.3 | $5.07 \cdot 10^{-5}$ | $2.23 \cdot 10^{-2}$ | $3.63 \cdot 10^{-21}$ | $1.42 \cdot 10^{-18}$ |
| A.4 | $6.00 \cdot 10^{-9}$ | $1.21 \cdot 10^{-3}$ | $1.37 \cdot 10^{-16}$ | $4.86 \cdot 10^{-7}$ |
| A.5 | $5.65 \cdot 10^{-9}$ | $1.29 \cdot 10^{-3}$ | $6.13 \cdot 10^{-17}$ | $5.18 \cdot 10^{-7}$ |
| A.6 | $2.33 \cdot 10^{-10}$ | $1.06 \cdot 10^{-3}$ | $1.50 \cdot 10^{-19}$ | $4.67 \cdot 10^{-7}$ |
| A.7 | $7.45 \cdot 10^{-8}$ | $1.07 \cdot 10^{-3}$ | $2.09 \cdot 10^{-19}$ | $1.66 \cdot 10^{-7}$ |
| A.8 | $1.86 \cdot 10^{-6}$ | $1.30 \cdot 10^{-3}$ | $9.19 \cdot 10^{-12}$ | $4.68 \cdot 10^{-7}$ |
| A.9 | $2.65 \cdot 10^{-6}$ | $1.29 \cdot 10^{-3}$ | $9.17 \cdot 10^{-12}$ | $5.18 \cdot 10^{-7}$ |

Table 11: Results in the Group B

| Subgroup | double | | quad | |
|---|---|---|---|---|
| | $max|\delta c_1|$ (m$^2$) | $max|S_1|$ (m$^2$) | $max|\delta c_1|$ (m$^2$) | $max|S_1|$ (m$^2$) |
| B.1 | $2.07 \cdot 10^{-6}$ | $9.92 \cdot 10^{-4}$ | $1.17 \cdot 10^{-11}$ | $5.07 \cdot 10^{-7}$ |
| B.2 | $1.87 \cdot 10^{-6}$ | $1.63 \cdot 10^{-3}$ | $6.04 \cdot 10^{-12}$ | $5.35 \cdot 10^{-7}$ |
| B.3 | $2.87 \cdot 10^{-5}$ | $2.14 \cdot 10^{-2}$ | $3.45 \cdot 10^{-21}$ | $1.53 \cdot 10^{-18}$ |
| B.4 | $8.45 \cdot 10^{-8}$ | $1.08 \cdot 10^{-3}$ | $1.85 \cdot 10^{-13}$ | $4.82 \cdot 10^{-7}$ |
| B.5 | $9.17 \cdot 10^{-8}$ | $1.44 \cdot 10^{-3}$ | $7.08 \cdot 10^{-16}$ | $5.22 \cdot 10^{-7}$ |
| B.6 | $6.62 \cdot 10^{-12}$ | $9.46 \cdot 10^{-4}$ | $1.50 \cdot 10^{-19}$ | $4.73 \cdot 10^{-7}$ |
| B.7 | $7.45 \cdot 10^{-8}$ | $1.14 \cdot 10^{-3}$ | $2.07 \cdot 10^{-19}$ | $1.66 \cdot 10^{-7}$ |
| B.8 | $2.16 \cdot 10^{-6}$ | $1.53 \cdot 10^{-3}$ | $9.19 \cdot 10^{-12}$ | $4.64 \cdot 10^{-7}$ |
| B.9 | $1.78 \cdot 10^{-6}$ | $1.50 \cdot 10^{-3}$ | $9.19 \cdot 10^{-12}$ | $5.13 \cdot 10^{-7}$ |

Table 12: Results in the Group C

| Subgroup | double | | quad | |
|---|---|---|---|---|
| | $max|\delta c_1|$ (m$^2$) | $max|S_1|$ (m$^2$) | $max|\delta c_1|$ (m$^2$) | $max|S_1|$ (m$^2$) |
| C.1 | $8.26 \cdot 10^{-8}$ | $1.11 \cdot 10^{-3}$ | $1.83 \cdot 10^{-13}$ | $4.98 \cdot 10^{-7}$ |
| C.2 | $8.15 \cdot 10^{-8}$ | $1.37 \cdot 10^{-3}$ | $1.39 \cdot 10^{-15}$ | $5.22 \cdot 10^{-7}$ |
| C.3 | $3.16 \cdot 10^{-7}$ | $3.83 \cdot 10^{-2}$ | $4.58 \cdot 10^{-23}$ | $1.57 \cdot 10^{-18}$ |
| C.4 | $5.44 \cdot 10^{-9}$ | $1.14 \cdot 10^{-3}$ | $1.33 \cdot 10^{-16}$ | $4.95 \cdot 10^{-7}$ |
| C.5 | $9.25 \cdot 10^{-8}$ | $1.13 \cdot 10^{-3}$ | $1.83 \cdot 10^{-13}$ | $4.93 \cdot 10^{-7}$ |
| C.6 | $5.70 \cdot 10^{-9}$ | $1.27 \cdot 10^{-3}$ | $5.95 \cdot 10^{-17}$ | $5.18 \cdot 10^{-7}$ |
| C.7 | $8.69 \cdot 10^{-8}$ | $1.45 \cdot 10^{-3}$ | $7.08 \cdot 10^{-16}$ | $5.22 \cdot 10^{-7}$ |
| C.8 | $1.25 \cdot 10^{-8}$ | $1.39 \cdot 10^{-3}$ | $9.28 \cdot 10^{-14}$ | $5.13 \cdot 10^{-7}$ |
| C.9 | $3.44 \cdot 10^{-24}$ | $1.10 \cdot 10^{-3}$ | $2.58 \cdot 10^{-53}$ | $5.11 \cdot 10^{-7}$ |

Table 13: Results in the Group D

| Subgroup | double | | quad | |
|---|---|---|---|---|
| | $max\|\delta c_1\|$ (m$^2$) | $max\|S_1\|$ (m$^2$) | $max\|\delta c_1\|$ (m$^2$) | $max\|S_1\|$ (m$^2$) |
| D.1 | $9.23 \cdot 10^{-8}$ | $1.08 \cdot 10^{-3}$ | $1.85 \cdot 10^{-13}$ | $5.13 \cdot 10^{-7}$ |
| D.2 | $5.88 \cdot 10^{-8}$ | $1.28 \cdot 10^{-3}$ | $6.97 \cdot 10^{-16}$ | $5.22 \cdot 10^{-7}$ |

Table 14: Results in the Group E

| double | | quad | |
|---|---|---|---|
| $max\|\delta c_1\|$ (m$^2$) | $max\|S_1\|$ (m$^2$) | $max\|\delta c_1\|$ (m$^2$) | $max\|S_1\|$ (m$^2$) |
| $2.30 \cdot 10^{-6}$ | $1.88 \cdot 10^{-3}$ | $1.06 \cdot 10^{-11}$ | $4.97 \cdot 10^{-7}$ |

From the results of Tables 10 – 14 we conclude that our method is almost independent of the qualitative characteristics of each geodesic (the order of magnitude of values remains almost the same).

In order to study in detail the differences between the double and quad precision, taking into account the precision of the input data, we searched for the geodesics of the whole set where the differences in the results at the end point were larger than $10^{-6}$ m for the $x_1, y_1, z_1$ and $10^{-10}$ deg for the $\beta_1, \lambda_1$ and $\alpha_1$. Only 436 geodesics were found, distributed in the subgroups as follows: 10 in B.5, 53 in C5, 11 in C.7, 1 in D.1 and 361 in D.2. In all these cases, one or both umbilical points are involved.

In addition, in order to examine the stability of the method and, furthermore, to obtain an estimate of the precision of the results, we modified the distance $s_{01}$ by $10^{-7}$ m in all groups of the dataset (except, of course, subgroup D2, where $s_{01}$ is fixed). After performing the new computation, we found that the differences in the values of ($x_1, y_1, z_1$) were always bounded by $10^{-7}$ m, therefore we conclude that 20000 steps in the numerical integration are adequate. With regard to the value of $\alpha_1$, most differences were of the order $10^{-10}$ deg or less, except for geodesics close to umbilical points, where the differences were up to $10^{-3}$ deg (subgroup B5). These expected differences are due to the rapid change of the $(\beta, \lambda)$ coordinate grid in the vicinity of the umbilical points and to the limited accuracy in the determination of $\beta_1, \lambda_1$ and, subsequently, of the vector **p**. In most cases, however, knowledge of the angle $\alpha_1$ and/or the Liouville constant is not required.

Finally, using only Cartesian coordinates in the algorithm, we computed vector **p** with Eqs. (63)-(65) and the Liouville constant with Eq. (81). Then, we repeated the computation of the data set and Table 15 presents some of the corresponding results. It is remarkable that, using Cartesian coordinates for the computation of vector **p**, we are led to a great loss of accuracy. This result is expected, since Eqs. (63)-(65) are numerically ill-behaved for a small difference between the $a_x$ and $a_y$ axes. In addition, in this case the computed vector **p** is a space vector not necessarily confined to the tangent plane of the triaxial ellipsoid.

Table 15: Results using Eqs. (63)-(65) and Eq. (81)

| Subgroup | quad | |
|---|---|---|
| | $max\|\delta c_1\|$ (m$^2$) | $max\|S_1\|$ (m$^2$) |
| A.1 | $5.79 \cdot 10^7$ | 12.6 |
| E | $1.70 \cdot 10^{-1}$ | $4.97 \cdot 10^{-7}$ |

## 7. Concluding remarks

A numerical solution of the geodesic initial value problem in Cartesian coordinates on a triaxial ellipsoid has been presented. The advantage of the proposed method is that it is a generalization of the method presented by Panou and Korakitis (2017) and hence can be used for a triaxial ellipsoid with arbitrary axes.

Comparing the results of using the vector **p** in ellipsoidal and Cartesian coordinates, we conclude that only expressing the vector **p** in ellipsoidal coordinates provides satisfactory results, i.e. works in the entire range of input data, it is stable and precise, so it is recommended for use. However, this requires the conversion of the Cartesian to ellipsoidal coordinates, therefore a new numerical method has been presented which is adequate to provide excellent results. In any case, it would be interesting to get a knowledge of the performance of other methods of conversion (e.g. Bektaş 2015).

**Acknowledgement:** The authors wish to thank Professor G. K. Aguirre, Perelman School of Medicine, University of Pennsylvania, for indicating to us his work "A model of the entrance pupil of the human eye" (https://github.com/gkaguirrelab/gkaModelEye), where the method presented here can be used. Also, we wish to thank Mr. O. Korakitis, Barcelona Supercomputer Center, for his assistance in exploiting the capabilities of the Quad-Precision Math Library.